\documentclass[conference]{IEEEtran}

\IEEEoverridecommandlockouts

\usepackage{cite}
\usepackage{overpic}
\usepackage{amsmath,amssymb,amsfonts}
\usepackage{graphicx}
\usepackage{textcomp}
\usepackage{xcolor}
\usepackage{float}
\usepackage{amsthm}
\usepackage{graphicx}
\usepackage{epstopdf}
\usepackage{subfigure}
\usepackage{adjustbox}
\usepackage{amsmath,bm}
\usepackage{amsfonts}
\usepackage{amssymb}
\usepackage{color}
\usepackage{multirow}
\usepackage{multicol}
\usepackage{soul,xcolor}
\usepackage{algorithm}
\usepackage{algpseudocode}%

\theoremstyle{plain}

\newcommand{\vect}[1]{\mathbf{#1}}

\def\diag{\mathrm{diag}}

\def\tr{\mathrm{tr}}

\def\Htran{\mbox{\tiny $\mathrm{H}$}}

\def\CN{\mathcal{N}_{\mathbb{C}}} 

\def\Htran{\mbox{\tiny $\mathrm{H}$}}

\def\CN{\mathcal{N}_{\mathbb{C}}} 

\begin{document}

\title{Cell-Free Massive MIMO with\\ Hardware-Impaired Wireless Fronthaul
\thanks{}
}

\author{\IEEEauthorblockN{ Özlem Tuğfe Demir\IEEEauthorrefmark{1} and Emil Björnson\IEEEauthorrefmark{2}\thanks{This work was carried out within the scope of the project 122C149 – Intelligent End-to-End Design of Energy-Efficient and Hardware Impairments-Aware Cell-Free Massive MIMO for Beyond 5G. \"O. T. Demir was supported by the 2232-B International Fellowship for Early Stage Researchers Programme funded by the Scientific and Technological Research Council of Türkiye (TÜBİTAK). E.~Bj\"ornson was supported by the Knut and Alice Wallenberg Foundation through a WAF grant.}}
\IEEEauthorblockA{ {$^*$Department of Electrical and Electronics Engineering, Bilkent University, Ankara, Turkiye
		}\\{$^\dagger$Department of Communication Systems, KTH Royal Institute of Technology, Stockholm, Sweden
		} 
   \\
		\IEEEauthorblockA{E-mail: \IEEEauthorrefmark{1}ozlemtugfedemir@bilkent.edu.tr, \IEEEauthorrefmark{2}emilbjo@kth.se}
}
}

\maketitle

\begin{abstract}
Cell-free massive MIMO (multiple-input–multiple-output) enhances spectral and energy efficiency compared to conventional cellular networks by enabling joint transmission and reception across a large number of distributed access points (APs). Since these APs are envisioned to be low-cost and densely deployed, hardware impairments—stemming from non-ideal radio-frequency (RF) chains—are unavoidable. While existing studies primarily address hardware impairments on the access side, the impact of hardware impairments on the wireless fronthaul link has remained largely unexplored.
In this work, we fill this important gap by introducing a novel amplify-and-forward (AF)-based wireless fronthauling scheme tailored for cell-free massive MIMO. Focusing on the uplink, we develop an analytical framework that jointly models the hardware impairments at both the APs and the fronthaul transceivers, derives the resulting end-to-end distorted signal expression, and quantifies the individual contribution of each impairment to the spectral efficiency. Furthermore, we design distortion-aware linear combiners that optimally mitigate these effects. Numerical results demonstrate significant performance gains from distortion-aware processing and illustrate the potential of the proposed AF fronthauling scheme as a cost-effective enabler for future cell-free architectures.
\end{abstract}

\section{Introduction}
Cell-free massive MIMO is a prominent 6G technology that relies on a large set of geographically distributed APs that collaboratively serve multiple UEs over shared time–frequency resources, thereby eliminating the need for traditional cell boundaries \cite{Ngo2017b,cell-free-book,ngo2024ultradense}. This high level of cooperation enhances service continuity and provides substantial gains in both spectral and energy efficiency compared to conventional cellular architectures. A key requirement for realizing these benefits is a reliable fronthaul connection between the APs and the central processing unit (CPU), which performs the joint uplink decoding and downlink signal generation. Indeed, the primary advantage of cell-free massive MIMO stems from its centralized processing capability \cite{cell-free-book}.

In the vast majority of studies on cell-free massive MIMO, the fronthaul links are assumed to have unlimited resolution and to operate over ideal, error-free wired connections. However, deploying such high-quality wired fronthaul—often relying on costly fiber infrastructure—is economically impractical for large-scale future cell-free networks. To overcome this limitation, several works have explored wireless fronthaul solutions for cell-free massive MIMO systems \cite{demirhan_wireless_fronthaul,ibrahim_wireless_fronthaul,ozan_wireless_fronthaul}. 
In these papers, the wireless fronthaul is modeled as a \emph{digital} link: the CPU and APs exchange user data symbols or quantized baseband samples over capacity-limited high-frequency fronthaul channels, corresponding to decode-and-forward–type or functional-split architectures rather than forwarding the received access signals in analog form \cite{demirhan_wireless_fronthaul,ibrahim_wireless_fronthaul,ozan_wireless_fronthaul}. 
In contrast, to the best of our knowledge, none of these studies have considered an \emph{amplify-and-forward} (AF) wireless fronthaul architecture, where each AP linearly precodes its received uplink signal and forwards it to the CPU for centralized processing. 
In our work, we introduce and analyze such an AF-type wireless fronthaul structure, in which the CPU performs the final multiuser data detection based on the jointly processed signals from all APs by  exploiting its large antenna array to spatially separate the different UE streams. 

In densely deployed cell-free massive MIMO systems, equipping APs with low-cost hardware is essential for scalability. Accurately modeling the resulting hardware impairments is therefore crucial to understanding and mitigating their detrimental effects. To the best of the authors’ knowledge, prior works have examined low-cost AP hardware primarily from the perspective of the access link \cite{impairment1,impairment2}. However, when APs also participate in wireless fronthaul transmission, they rely on similarly cost-efficient hardware for the uplink fronthaul, which introduces additional distortion. Although some studies consider limited fronthaul resolution \cite{impairment3,impairment4}, none of them explicitly account for hardware impairments on the wireless fronthaul chain. In this work, we address this gap by jointly modeling both access-side and fronthaul-side impairments.

In this work, we address two fundamental research questions that have
remained open in the context of cell-free massive MIMO with wireless
fronthaul:

\textbf{(Q1) How does an amplify-and-forward (AF) wireless fronthaul
architecture perform compared to the ideal centralized operation with
perfect wired fronthaul?}

While AF relaying is widely used in cooperative communication, its
implications for cell-free massive MIMO have not been analytically
characterized. Unlike the conventional centralized benchmark—where the
CPU operates directly on uplink signals—the AF architecture forwards
analog signals that include propagation noise and front-end distortions.
Thus, a rigorous comparison between AF-based wireless fronthauling and
the ideal centralized architecture is still missing.

\textbf{(Q2) How do hardware impairments at both the access and wireless
fronthaul chains jointly affect the end-to-end performance of AF-based
cell-free massive MIMO?}

Existing studies primarily consider impairments only on the access link
or assume digital fronthaul with quantized samples. However, in AF
wireless fronthauling, the distortions introduced by low-cost hardware
at the APs and fronthaul transceivers accumulate and propagate through
both hops before reaching the CPU. Understanding this joint distortion
propagation is essential for designing receivers that can operate
reliably under practical hardware constraints.

To answer Q1 and Q2, we develop the first analytical framework that
models AF-based wireless fronthaul together with hardware impairments on
both links, derives the resulting end-to-end distorted signal
expression, and enables distortion-aware linear combining at the CPU.

\section{System Model}

We consider a cell-free massive MIMO system with $L$ APs, each with $N$ antennas for uplink reception 
and wireless fronthaul transmission. The CPU has $M$ antennas to receive the wireless fronthaul signals from the APs. The channel from UE $k$ to AP $l$ is $\mathbf{h}_{kl} \in \mathbb{C}^N$, 
and the fronthaul channel from AP $l$ to the CPU is $\mathbf{G}_l \in \mathbb{C}^{M \times N}$. We assume that the uplink access transmission and the wireless fronthaul transmission occur in separate time slots. In this work, we account for hardware impairments on both the access side and the fronthaul side—an aspect that has been largely overlooked in prior studies. A further key novelty of our approach is the use of an amplify-and-forward scheme for the wireless fronthaul transmission, which has not been explored in this context.

We assume the availability of perfect channel state information (CSI) at both the local APs and the central CPU. The consideration of imperfect CSI and its acquisition at the CPU through the proposed amplify-and-forward wireless fronthauling scheme is left for future work.

\subsection{Access-Side Hardware Impairments}
In the first slot of the amplify-and-forward wireless fronthauling scheme, the UEs send their data signals to the APs. We denote the unit-power symbol of UE $k$ by $s_k$, i.e., $\mathbb{E}\{|s_k|^2\}=1$ and and its uplink transmit power as $p_k>0$.
The received signal at AP $l$ is modeled as
\begin{align}
\mathbf{y}_l
= \sqrt{\kappa_{\mathrm{ac}}}
\sum_{k=1}^K \mathbf{h}_{kl}\sqrt{p_k} s_k
+ \boldsymbol{\eta}_{\mathrm{ac},l}
+ \mathbf{n}_{\mathrm{ac},l}, \label{eq:access}
\end{align}
where $\kappa_{\mathrm{ac}} \in (0,1]$ denotes access-side hardware quality factor, with larger values indicating weaker hardware impairment effects.
The distortion $\boldsymbol{\eta}_{\mathrm{ac},l}$ is modeled as
\[
\boldsymbol{\eta}_{\mathrm{ac},l} \sim
\mathcal{CN}(\mathbf{0}, \mathbf{D}_{\mathrm{ac},l}),
\]
with a covariance matrix proportional to the signal power, given as \cite{massive_mimo_book}
\begin{align}
    \vect{D}_{\mathrm{ac},l} = \diag\left( (1-\kappa_{\mathrm{ac}}) \sum_{k=1}^Kp_k\vect{h}_{kl}\vect{h}_{kl}^{\Htran}\right),
\end{align}
where $\diag(\cdot)$ extracts the diagonal part of the matrix by setting off-diagonal entries to zero. The distortion is independent of the other signals and also spatially white. The additive independent noise is denoted by $\vect{n}_{\mathrm{ac},l}\sim \CN(\vect{0},\sigma^2\vect{I}_N)$.

\subsection{Wireless Fronthaul Transmission with Impairments}

In the second slot, each AP $l$ applies a fronthaul precoder $\mathbf{P}_l\in \mathbb{C}^{N\times N}$
(with the gain factor implicitly absorbed into $\mathbf{P}_l$ for notational convenience). The impaired transmitted signal from AP $l$ is given as
\begin{align}
\tilde{\mathbf{y}}_l
= \sqrt{\kappa_{\mathrm{frt}}}\,\mathbf{P}_l \mathbf{y}_l
+ \boldsymbol{\eta}_{\mathrm{frt},l}, \label{eq:transmit-fronthaul}
\end{align}
where $\kappa_{\mathrm{frt}}\in (0,1]$ is the fronthaul hardware quality factor and
$\boldsymbol{\eta}_{\mathrm{frt},l}$ is fronthaul distortion modeled as
$
\boldsymbol{\eta}_{\mathrm{frt},l}
\sim \mathcal{CN}(\mathbf{0}, \mathbf{D}_{\mathrm{frt},l})
$. The covariance matrix of the distortion is given as
    \begin{align}
    \vect{D}_{\mathrm{frt},l} = \diag\left( (1-\kappa_{\mathrm{frt}})\left( \sum_{k=1}^Kp_k\vect{P}_l\vect{h}_{kl}\vect{h}_{kl}^{\Htran}\vect{P}_l^{\Htran}+\sigma^2\vect{P}_l\vect{P}_l^{\Htran}\right)\right).
\end{align}
The CPU receives
\begin{align}
\mathbf{y}
= \sum_{l=1}^L \mathbf{G}_l \tilde{\mathbf{y}}_l 
+ \mathbf{n}_{\mathrm{frt}}, \label{eq:cpu_received}
\end{align}
where $\mathbf{n}_{\mathrm{frt}}\sim\CN(\vect{0},\sigma^2\vect{I}_M)$ is the  additive independent noise. Next, we will consider the linear combining at the CPU to obtain soft estimates of each UE symbol.  

\section{Linear Combining at the CPU}
This section derives the expression of the received signal that is
processed by the linear receive combiner at the CPU for detecting each
UE’s uplink symbol from the AF-forwarded fronthaul transmissions. 
Substituting \eqref{eq:access} into \eqref{eq:transmit-fronthaul} and then into \eqref{eq:cpu_received}, we obtain the received signal at the CPU as
\begin{align}
\mathbf{y}
=& \underbrace{\sqrt{\kappa_{\mathrm{frt}}\kappa_{\mathrm{ac}}}
\sum_{k=1}^K \sqrt{p_k}
\sum_{l=1}^L \mathbf{G}_l \mathbf{P}_l \mathbf{h}_{kl} s_k}_{\text{desired + interference}} \nonumber\\
& + \underbrace{
\sqrt{\kappa_{\mathrm{frt}}}
\sum_{l=1}^L \mathbf{G}_l \mathbf{P}_l
(\boldsymbol{\eta}_{\mathrm{ac},l} + \mathbf{n}_{\mathrm{ac},l})
}_{\text{amplified access distortion \& noise}} \nonumber\\
& + \underbrace{
\sum_{l=1}^L \mathbf{G}_l \boldsymbol{\eta}_{\mathrm{frt},l}
}_{\text{fronthaul distortion}}
+ \mathbf{n}_{\mathrm{frt}}. \label{eq:decomposed}
\end{align}
For notational convenience, we define the effective channel
\begin{align}
\mathbf{b}_{kl} = \mathbf{G}_l \mathbf{P}_l \mathbf{h}_{kl}.
\end{align}
We let $\mathbf{v}_k$ be the receive combining vector for UE $k$ at the CPU. 
The CPU forms the soft estimate
\begin{align}
\hat{s}_k = \mathbf{v}_k^{\Htran} \mathbf{y}.
\end{align}
Using \eqref{eq:decomposed}, this becomes
\begin{align}
\hat{s}_k
=& \sqrt{\kappa_{\mathrm{frt}}\kappa_{\mathrm{ac}}}\sqrt{p_k}
\sum_{l=1}^L \mathbf{v}_k^{\Htran} \mathbf{b}_{kl} s_k \nonumber\\
&+ \sqrt{\kappa_{\mathrm{frt}}\kappa_{\mathrm{ac}}}
\sum_{i\neq k} \sqrt{p_i}\sum_{l=1}^L
\mathbf{v}_k^{\Htran} \mathbf{b}_{il} s_i \nonumber\\
&+ \sqrt{\kappa_{\mathrm{frt}}}
\sum_{l=1}^L \mathbf{v}_k^{\Htran}
\mathbf{G}_l \mathbf{P}_l (\boldsymbol{\eta}_{\mathrm{ac},l} + \mathbf{n}_{\mathrm{ac},l}) \nonumber\\
&+ \sum_{l=1}^L \mathbf{v}_k^{\Htran}\mathbf{G}_l\boldsymbol{\eta}_{\mathrm{frt},l}
+ \mathbf{v}_k^{\Htran} \mathbf{n}_{\mathrm{frt}}.
\end{align}

\section{Spectral Efficiency with Optimal Linear Combiner}

Using standard methodology for linear receivers in systems with uncorrelated additive distortions, the SE of UE $k$ is
\begin{align}
\mathrm{SE}_k = \log_2\left(1+\mathrm{SINR}_k\right)
\end{align}
where the effective SINR is 
\begin{align}
\mathrm{SINR}_k
= \frac{
\kappa_{\mathrm{frt}}\kappa_{\mathrm{ac}} p_k
\left\lvert \sum_{l=1}^L \mathbf{v}_k^{\Htran} \mathbf{b}_{kl} \right\rvert^2
}{
I_k + D_{\mathrm{ac},k} + D_{\mathrm{frt},k} + \sigma^2 \|\mathbf{v}_k\|^2
},
\label{eq:sinr_basic}
\end{align}
where each term corresponds to:
\begin{itemize}
\item \textbf{Multiuser interference:}
\begin{align}
I_k = \kappa_{\mathrm{frt}}\kappa_{\mathrm{ac}}
\sum_{i\neq k} p_i
\left\lvert \sum_{l=1}^L \mathbf{v}_k^{\Htran} \mathbf{b}_{il} \right\rvert^2.
\end{align}

\item \textbf{Access distortion term:}
\begin{align}
D_{\mathrm{ac},k}
= \kappa_{\mathrm{frt}}
\sum_{l=1}^L
\mathbf{v}_k^{\Htran} \mathbf{G}_l \mathbf{P}_l (\mathbf{D}_{\mathrm{ac},l}+\sigma^2\vect{I}_N)
\mathbf{P}_l^{\Htran} \mathbf{G}_l^{\Htran} \mathbf{v}_k.
\end{align}

\item \textbf{Fronthaul distortion term:}
\begin{align}
D_{\mathrm{frt},k}
= \sum_{l=1}^L
\mathbf{v}_k^{\Htran} \mathbf{G}_l \mathbf{D}_{\mathrm{frt},l}
\mathbf{G}_l^{\Htran} \mathbf{v}_k.
\end{align}
\end{itemize}
By defining the interference-plus-distortion covariance matrix
\begin{align}
\mathbf{R}_k &=
\kappa_{\mathrm{frt}}\kappa_{\mathrm{ac}}
\sum_{i\neq k} p_i
\left(\sum_{l=1}^L \mathbf{b}_{il}\right)\left(\sum_{l'=1}^L\mathbf{b}_{il'}^{\Htran}\right) \nonumber \\
&\quad+ \kappa_{\mathrm{frt}}\sum_{l=1}^L
\mathbf{G}_l \mathbf{P}_l (\mathbf{D}_{\mathrm{ac},l}+\sigma^2\vect{I}_N) \mathbf{P}_l^{\Htran} \mathbf{G}_l^{\Htran} \nonumber \\
&\quad+ \sum_{l=1}^L \mathbf{G}_l \mathbf{D}_{\mathrm{frt},l} \mathbf{G}_l^{\Htran}
+ \sigma^2 \mathbf{I}_M,
\end{align}
the SINR can be written as
\begin{align}
    \mathrm{SINR}_k = \frac{
\kappa_{\mathrm{frt}}\kappa_{\mathrm{ac}} p_k
\left\lvert \mathbf{v}_k^{\Htran}\sum_{l=1}^L  \mathbf{b}_{kl} \right\rvert^2
}{\vect{v}_k^{\Htran}\vect{R}_k\vect{v}_k}.
\end{align}
Since this is a generalized Rayleigh quotient, the SINR-optimal distortion-aware linear combiner can be obtained as
\begin{align}
\mathbf{v}_k^{\mathrm{opt}}
\propto \mathbf{R}_k^{-1}
\sum_{l=1}^L \mathbf{b}_{kl}.
\end{align}
This leads to the maximum SINR
\begin{align}
    \mathrm{SINR}_k^{\star} = \kappa_{\mathrm{frt}}\kappa_{\mathrm{ac}}p_k\left(\sum_{l=1}^L \mathbf{b}_{kl}\right)^{\Htran}\vect{R}_k^{-1}\left(\sum_{l=1}^L \mathbf{b}_{kl}\right).
\end{align}
\section{Fronthaul Precoding Structure}

In this section, we will go through two wireless fronthaul precoding choices. To this end, we decompose the precoding matrix as
\begin{align}
    \vect{P}_l = \alpha_l \overline{\vect{P}}_l,
\end{align}
where $\alpha_l$ is the scaling factor that adjusts the gain of the precoding matrix to have a certain transmit power at the APs. The transmit power of AP $l$ is computed as
\begin{align}
    p_{\mathrm{frt},l} &= \tr\left( \sum_{k=1}^Kp_k\vect{P}_l\vect{h}_{kl}\vect{h}_{kl}^{\Htran}\vect{P}_l^{\Htran}+\sigma^2\vect{P}_l\vect{P}_l^{\Htran}\right) \nonumber\\
    &=\alpha_l^2 \tr\left( \sum_{k=1}^Kp_k\overline{\vect{P}}_l\vect{h}_{kl}\vect{h}_{kl}^{\Htran}\overline{\vect{P}}_l^{\Htran}+\sigma^2\overline{\vect{P}}_l\overline{\vect{P}}_l^{\Htran}\right).
\end{align}
After fixing $\overline{\vect{P}}_l$, we can scale $\alpha_l$ to have a desired transmit power level $p_{\mathrm{frt},l}$ as
\begin{align}
    \alpha_l = \frac{\sqrt{p_{\mathrm{frt},l}}}{\sqrt{\tr\left( \sum_{k=1}^Kp_k\overline{\vect{P}}_l\vect{h}_{kl}\vect{h}_{kl}^{\Htran}\overline{\vect{P}}_l^{\Htran}+\sigma^2\overline{\vect{P}}_l\overline{\vect{P}}_l^{\Htran}\right)}}.
\end{align}

The first choice is the \emph{identity precoding}, i.e., $\overline{\vect{P}}_l = \mathbf{I}_N$, and this provides a baseline without spatial shaping.

A potentially more effective strategy is inspired by the optimal SVD precoder for point-to-point MIMO transmission. By matching the left and right singular spaces of the precoding matrix to the access and fronthaul channels, respectively, we obtain the scheme we refer to as \emph{bi-SVD precoding}. To this end, we denote the SVD of the matrix $\vect{G}_l$ and the matrix $\vect{H}_l=[\vect{h}_{1l} \ \ldots \ \vect{h}_{Kl}]$ as
\begin{align}
\mathbf{G}_l = \mathbf{U}_{G,l} \boldsymbol{\Sigma}_{G,l} \mathbf{V}_{G,l}^{\Htran},\qquad
\mathbf{H}_l = \mathbf{U}_{H,l} \boldsymbol{\Sigma}_{H,l} \mathbf{V}_{H,l}^{\Htran}.
\end{align}
Then the bi-SVD precoder is given as
\begin{align}
\overline{\mathbf{P}}_l = \mathbf{V}_{G,l}\mathbf{U}_{H,l}^{\Htran}
\end{align}
which aligns the access link and fronthaul link, improving conditioning
of the effective channels $\mathbf{b}_{kl}$.

\section{Numerical Results}
In this section, we quantify the performance of the proposed AF-relaying-based wireless fronthauling scheme and compare it with the ideal centralized operation under perfect hardware conditions. We also assess the performance degradation caused by hardware impairments at both the access and fronthaul sides. In the figure legends, the ideal centralized architecture with perfect hardware is labeled as “Perfect (centralized)”, while the proposed AF-based wireless fronthaul scheme under perfect hardware is denoted “Perfect (wireless)”. The remaining cases incorporate hardware impairments and are evaluated under two different strategies. The first strategy applies the distortion-aware combining derived in this paper. The second strategy computes the receive combining vector by assuming perfect hardware (i.e., $\kappa_{\rm ac}=\kappa_{\rm frt}=1$), but its performance is still evaluated under the actual impaired hardware, identical to the distortion-aware case. This allows us to clearly illustrate the benefit of using hardware-adaptive combining.

We adopt the path loss expressions and correlated shadowing model from \cite{cell-free-book}, and generate spatial correlation matrices following that methodology for a sub-6\,GHz channel with a $15^\circ$ angular standard deviation in the local scattering model. Differing slightly from the model in \cite{cell-free-book}, we consider a bandwidth of $50$\,MHz and set the noise variance to $\sigma^2 = -94$\,dBm (corresponding to a 3\,dB noise figure). In each random setup, $16$ APs are uniformly and independently deployed in an $800\times 800$\,m$^2$ square area. Each AP is equipped with $N=4$ antennas, and there are $K=8$ UEs uniformly and independently dropped in the same area. The AP–UE channels follow correlated Rayleigh fading, while the AP–CPU channels follow correlated Rician fading with a K-factor of 10\,dB. The APs and CPU are positioned 10\,m above the UEs. Unless otherwise stated, the hardware quality factors are set to $\kappa_{\rm ac} = \kappa_{\rm frt} = 0.9$. Each UE transmits with power $p_k = 0.1$\,W, $\forall k$, and each AP scales its fronthaul transmit precoder to satisfy a total transmit power of $10$\,W.

\begin{figure}[t] 
    \centering
        \includegraphics[width=0.5\textwidth, trim=0.2cm 0cm 1cm 0.2cm, clip]{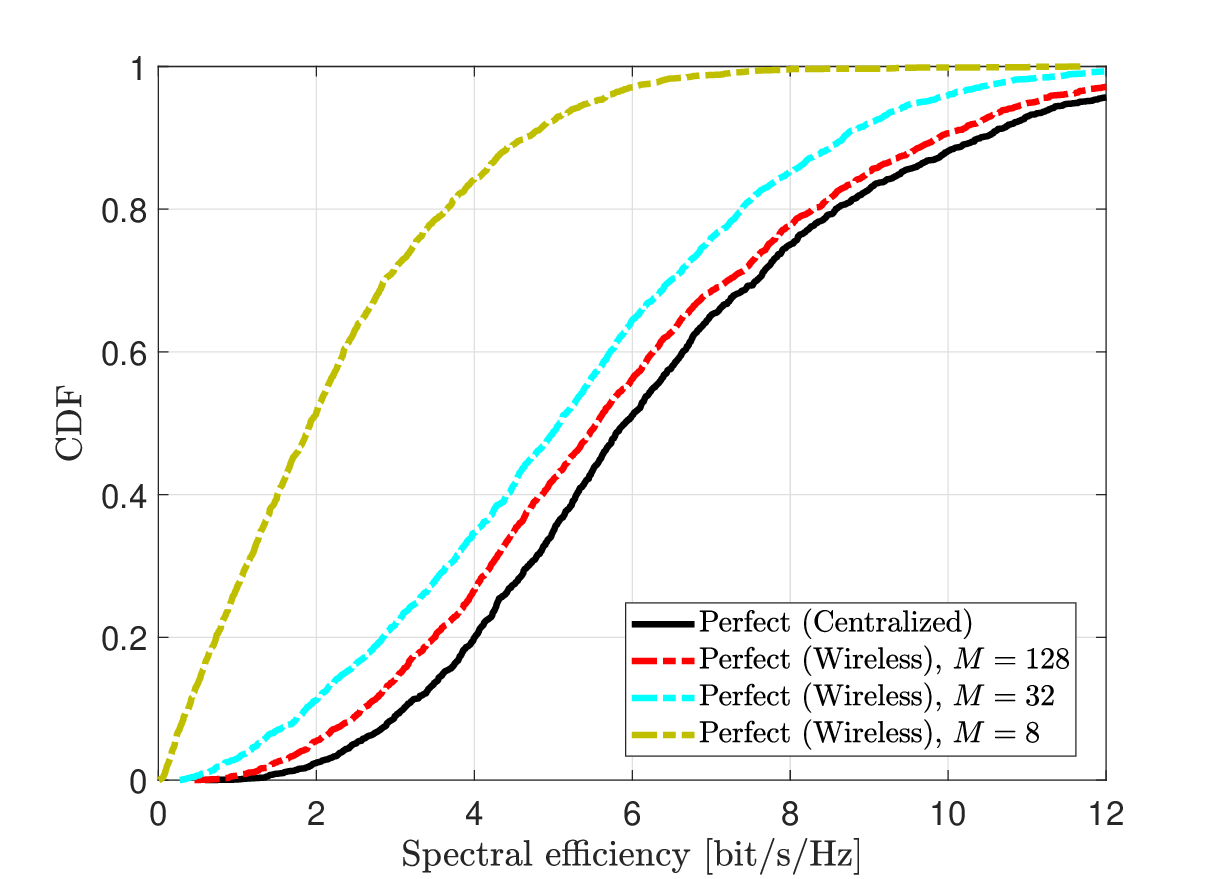} 
        \caption{The CDF of the uplink SE for centralized processing with ideal wired fronthaul and the proposed AF-based wireless fronthaul scheme under perfect hardware. Results are shown for different numbers of CPU antennas $M$.}
        \label{fig:1}
 \end{figure}

In Fig.~\ref{fig:1}, the cumulative distribution function (CDF) of the SE per UE is plotted. We consider the perfect-hardware case and benchmark the proposed AF-relaying-based wireless fronthauling architecture against the ideal centralized operation of a cell-free massive MIMO system employing bi-SVD precoding at the APs. We investigate several different values of the number of CPU antennas $M$. The key observation is that when $M$ is small, the performance of wireless fronthauling is significantly below that of the ideal centralized solution. However, as $M$ increases, the performance gap narrows, and the wireless-fronthaul architecture approaches the centralized benchmark with wired fronthaul. An important future research direction is to investigate the extent to which optimized fronthaul precoding and transmit power control can further narrow this gap.

\begin{figure}[t] 
    \centering
        \includegraphics[width=0.5\textwidth, trim=0.2cm 0cm 1cm 0.2cm, clip]{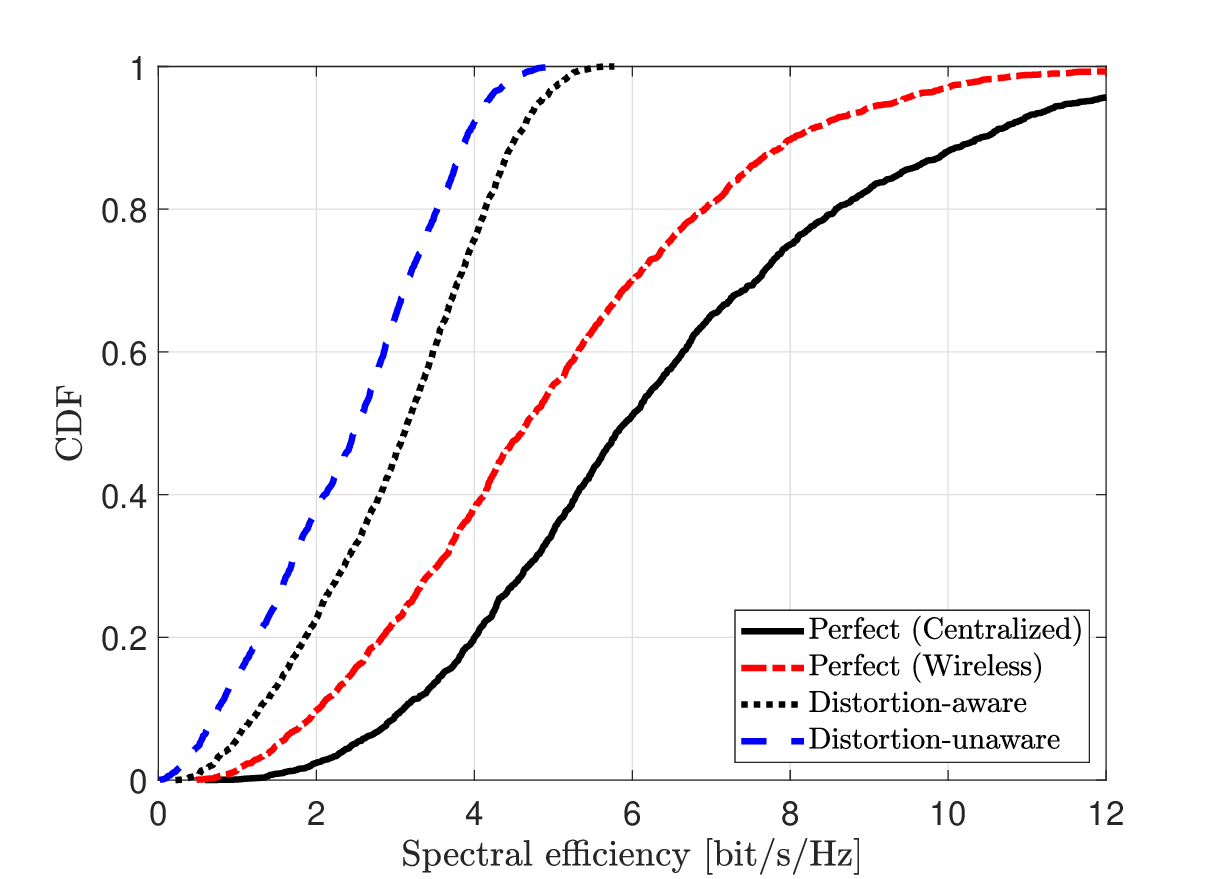} 
        \caption{The CDF of SE for identity fronthaul precoding with 
$M=128$ CPU antennas. Both distortion-aware and distortion-unaware combining strategies are compared against the ideal centralized benchmark.}
        \label{fig:2}
 \end{figure}

\begin{figure}[t] 
    \centering
        \includegraphics[width=0.5\textwidth, trim=0.2cm 0cm 1cm 0.2cm, clip]{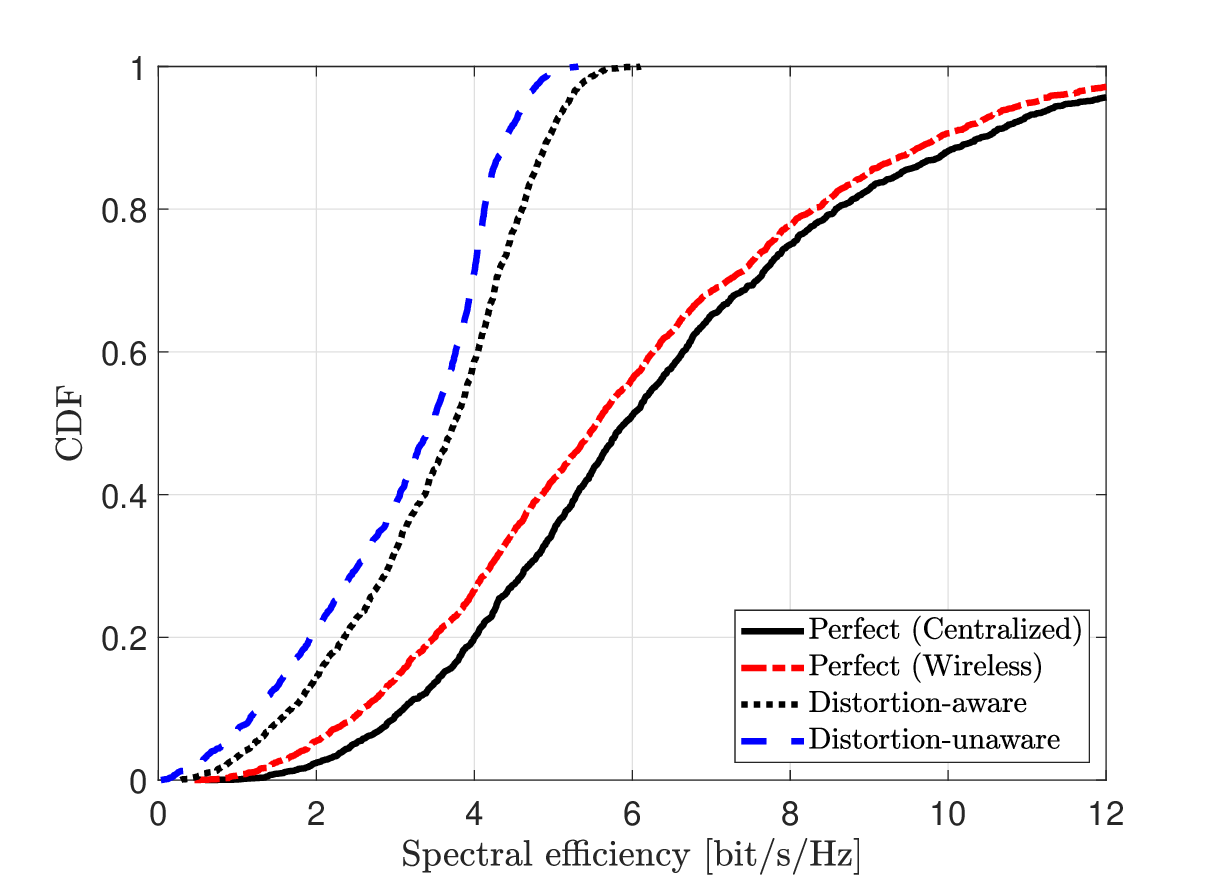} 
        \caption{The CDF of SE for bi-SVD fronthaul precoding with $M=128$ CPU antennas. The improved conditioning offered by bi-SVD results in a right-shifted SE distribution relative to identity precoding.}
        \label{fig:3}
 \end{figure}

In Figs.~\ref{fig:2} and \ref{fig:3}, we evaluate the SE performance using identity precoding and bi-SVD precoding for fronthaul transmission, respectively, with $M = 128$ CPU antennas. As observed from the figures, the curves obtained with bi-SVD precoding shift to the right, indicating superior performance. This improvement stems from the ability of bi-SVD precoding to better condition the effective channel by aligning the relevant eigenspaces, making it a more favorable choice compared to spatially agnostic identity precoding.

Furthermore, both the access- and fronthaul-side distortions significantly degrade the SE, particularly for UEs that would otherwise achieve high rates. When identity precoding is used, incorporating distortion-aware combining becomes even more critical. Nevertheless, for both precoding strategies, distortion-aware processing consistently improves the SE, with the largest gains observed for high-rate UEs. 

\begin{figure}[t] 
    \centering
        \includegraphics[width=0.5\textwidth, trim=0.2cm 0cm 1cm 0.2cm, clip]{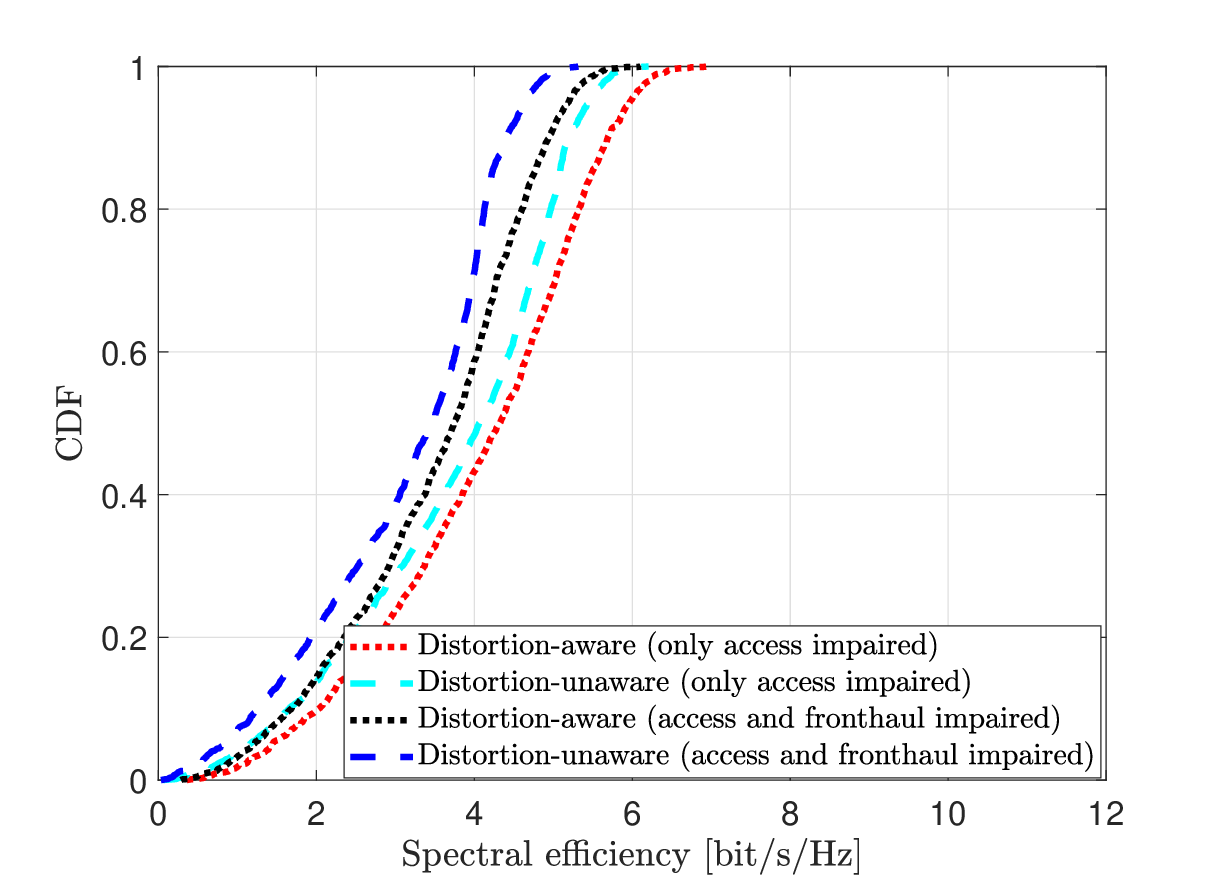} 
        \caption{The CDF of SE comparing two hardware impairment scenarios: (i) only access-side impairments and (ii) joint access- and fronthaul-side impairments. Distortion-aware and distortion-unaware combining are evaluated to highlight the relative performance loss.}
        \label{fig:4}
 \end{figure}

In Fig.~\ref{fig:4}, we compare two scenarios: (i) only access-side impairments, with $\kappa_{\rm frt}=1$ and $\kappa_{\rm ac}=0.9$, and (ii) both access- and fronthaul-side impairments, with $\kappa_{\rm frt}=\kappa_{\rm ac}=0.9$. As illustrated in the figure, when the fronthaul is also impaired, even distortion-aware combining cannot surpass the performance of the distortion-unaware scheme in the case with no fronthaul impairments. For both scenarios, distortion-aware processing yields higher SE, particularly for high-rate UEs. However, for mid-rate UEs, distortion-awareness becomes more critical when both access and fronthaul chains are impaired, compared to the case where only the access side is impaired.

\section{Conclusion}
In this paper, we introduced a novel amplify-and-forward wireless fronthaul architecture for cell-free massive MIMO and developed the first analytical framework that jointly incorporates hardware impairments at both the access and fronthaul chains. By decomposing the end-to-end received signal, we quantified how distortions propagate through the system and derived a distortion-aware linear combiner that effectively mitigates these effects.

Our numerical results reveal several important insights. First, wireless fronthaul with low-cost hardware can closely approach the performance of ideal wired fronthaul when the CPU is equipped with a sufficient number of antennas and appropriate fronthaul precoding—such as the proposed bi-SVD design—is employed. Second, hardware impairments at both stages significantly degrade performance, especially for high-rate UEs, underscoring the importance of incorporating distortion-aware processing. 

Overall, the proposed AF-based wireless fronthaul framework provides a low-cost and scalable alternative to traditional wired fronthaul, with substantial performance gains enabled by distortion-aware combining and channel-aware fronthaul precoding. Future work includes extending the analysis to imperfect CSI acquisition and optimizing fronthaul resource allocation jointly with access-side signal processing.

\bibliographystyle{IEEEtran}
\bibliography{IEEEabrv,refs}

\end{document}